\documentclass[english]{nature}
\makeatletter

\usepackage{amsmath,amsfonts,amssymb}
\usepackage{graphicx}

\newcommand{\mean}[1]{\langle#1\rangle}
\def\ket#1{\left| #1\right>}

\newcommand{\ham}{\mathcal{H}}

\begin{document}
\title{High-sensitivity diamond magnetometer with nanoscale resolution} 
\author{J. M. Taylor$^{1*}$, P. Cappellaro$^{2,3*}$,
L. Childress$^{2,4}$, L. Jiang$^2$, D.Budker$^5$, P. R. Hemmer$^6$, A.Yacoby$^2$, R. Walsworth$^{2,3}$, M. D. Lukin$^{2,3}$}
\maketitle

\begin{affiliations}
\item Department of Physics, Massachusetts Institute of Technology, Cambridge, MA 02139  USA
\item Department of Physics, Harvard University, Cambridge, MA 02138  USA  
\item Harvard-Smithsonian Center for Astrophysics, Cambridge, MA 02138
USA
\item Department of Physics, Bates College,  Lewiston, ME  04240 USA
\item Department of Physics, University of California, Berkeley, CA 94720, USA  
\item Department of Electrical and Computer Engineering, Texas A\&M University, College Station, TX 77843 USA
\end{affiliations}

\begin{abstract}
We present a novel approach to the detection of weak magnetic fields that takes advantage of
recently developed techniques for the coherent control of solid-state
electron spin quantum bits. Specifically, we investigate a magnetic sensor based on Nitrogen-Vacancy centers in room-temperature diamond.  
We discuss two important applications of this technique: a nanoscale magnetometer that could potentially detect precession of single nuclear spins  and
an optical magnetic field imager combining spatial resolution ranging from micrometers to millimeters with a sensitivity approaching few femtotesla/Hz$^{1/2}$.
\end{abstract}

The detection of weak magnetic fields with high spatial resolution
is an important problem in diverse areas ranging from fundamental
physics and material science to data storage and biomedical science.
Over the past few decades, a wide variety of magnetic sensors have
been developed using approaches including superconducting quantum
interference devices (SQUIDs)\cite{bending99}, the Hall effect in semiconductors\cite{chang:Hallmag}, atomic
vapor and BEC-based magnetometry\cite{budker02,auzinsh04,savukov05, vengalattoreBECmagnetometer,zhaoBECmagnetometer}, and magnetic resonance force
microscopy\cite{mamin07,seton97,schlenga99}. 
In this article we present a novel approach to high spatial resolution magnetic field detection, 
using systems currently explored as quantum bits:
isolated electronic spins in a solid.  We focus on
spins associated with Nitrogen-Vacancy (NV) color centers in
diamond\cite{jelezko:076401} (Fig. 1 a-b), since they can be individually addressed, optically
polarized and detected, and exhibit excellent coherence properties
even at room temperature\cite{jelezko:130501,hanson06,childress06}.
Recently, coherent control of NV electronic spin qubits has been used
to sense and manipulate nearby individual
electronic\cite{epstein05,wracthnphys} and nuclear spins\cite{dutt06}
in a diamond lattice.  Here we describe how such a system can also be
used for the precision sensing and imaging of external magnetic fields.  

We discuss two types of potential implementations of such sensors.
First, a single sensing spin  
confined in a nanoscale region can be brought
in direct proximity to a magnetic field source, such as
an electron or nuclear spin.  For example, a diamond nanocrystal (10-50 nm in size)
containing a single NV center can be attached to a tip of a
scanning probe (Fig. 1c)\cite{Kuhn01}.  
Second, a bulk diamond
sample with a high density of NV centers can be used to sense fields created by
remote objects with ultra-high sensitivity and sub-$\mu$m spatial resolution (Fig. 1d).

\section*{Magnetometry with single electronic spin qubits}

The operating principles of our approach are closely related to
those of magnetometers based on spin precession in atomic vapors. In particular, detecting the relative energy shift induced
by a magnetic field $b$ between two Zeeman sublevels allows for a precise
determination of an applied DC or AC magnetic field.  Ultimately,
sensitivity is determined by the spin
coherence time and by the spin projection-noise. 
Although solid-state electronic spins have shorter coherence times
than gaseous atoms, quantum control techniques can decouple them from
the local environment and from each other, as we show below, leading
to a substantial improvement in their sensitivity to external,
time-varying magnetic fields, while retaining the desirable features of a robust solid sensor.

The canonical approach to detecting a Zeeman shift uses a
Ramsey-type sequence as illustrated in Fig.~\ref{f:sens}a.  A
$\pi/2$-pulse creates a superposition of two Zeeman levels, which
acquire a relative phase $\phi = \delta \omega\ \tau\propto \frac{g
\mu_B}{\hbar} b \tau$ from the external field $b$ during the free
evolution interval $\tau$ (here $\mu_B$ is the Bohr magneton and
$g\approx 2$ for NV centers).  Another $\pi/2$-pulse transforms the
relative phase into a population difference, which is measured optically and from which the Zeeman shift is
inferred.  For small $\phi$, the magnetometer signal $\mathcal{S}$ (proportional to the induced population difference) depends linearly
on the magnetic field: $\mathcal{S}\approx \frac {g \mu_B}{\hbar}b
\tau$. During the total averaging interval $T$, $T/\tau$ measurements
can be made, yielding a shot-noise-limited sensitivity $\eta$ given by
the minimum detectable field,
$b_{min}\equiv\eta/\sqrt{T}=\frac{\hbar}{g \mu_B} \frac 1{\sqrt{\tau
  T}}$.

Increasing the interrogation time $\tau$ improves the sensitivity
until random (environmental) perturbations lead to decay of the free-precession signal.  
In the case of solid-state spin systems, the coherence is limited by
interactions with nearby lattice nuclei and paramagnetic impurities, resulting in an ensemble dephasing
time $T_2^*$.  Furthermore, there will be a finite number of fluorescence 
photons collected and detected, leading to additional photon shot noise, and a finite contrast to the Ramsey fringes.  
We describe these effects by a single parameter $C \leq 1$, which approaches unity for ideal, single-shot readout (see Methods).  The optimum sensitivity of a 
magnetometer based on a single electronic spin, achieved for $\tau \sim T_2^*$, is given by
\begin{equation}
\eta_{DC}
\approx \frac{\hbar}{g \mu_B C \sqrt{T_2^*}}\ .
\label{etaDC}
\end{equation}
For current experiments\cite{childress06}, with detection efficiency $\sim 10^{-3}$, $C \approx 0.05$ and $T_2^* \sim 1\ \mu$s.  This yields an optimal sensitivity $\sim 1\ \mu$T/Hz$^{1/2}$.  
Improving the collection efficiency to $\eta \sim 5\%$ yields $C\approx 0.3$ and leads to a sensitivity $\sim 120$ nT/Hz$^{1/2}$. 

Coherent control techniques can improve the sensitivity for AC fields.
Due to the long correlation times characteristic of dipolar interactions between spins in systems such as diamond---the principal source of dephasing---spin echo techniques can dramatically extend the coherence time.  
Specifically, by adding an additional microwave $\pi$ pulse to
the Ramsey sequence at time $\tau/2$, the Hahn echo sequence (Fig. 2a)
removes the effect of environmental perturbations whose correlation time is long compared to $\tau$.  Thus a signal field $b(t)$
oscillating in-phase with the pulse sequence produces an overall additive
phase shift, leading to a total phase accumulation, $\delta \phi =
\frac{g \mu_B}{\hbar} [\int_0^{\tau/2} b(t) dt- \int_{\tau/2}^\tau
b(t) dt]$.  For a signal field $b(t) =b \sin(\nu t + \varphi_0)$, this
yields $\delta \phi= \frac{g \mu_B}{\hbar} b\tau f(\nu
\tau,\varphi_0)$, with $f(x,\varphi_0) = \frac{\sin^2(x/4) \cos(x/2 +
\varphi_0)}{x/4}$.  In essence, the spin echo allows us to extend
the interrogation time $\tau$ from the limit set by $T_2^*$ up to
a value $T_2$ which is close to the {\em intrinsic} spin coherence time, at the cost
of a reduced bandwidth and insensitivity to frequencies $\lesssim
1/T_2$. For maximal response to CW signals with known
frequency and phase (assuming small $b$), we find $\tau= 2 \pi/\nu$
and $\varphi_0 = 0$ to be optimal.  For signals whose time dependence is a-priori unknown, it is useful to measure the signal variance, which provides equivalent performance (see Methods).
In either case, the sensitivity is improved by $\sim \sqrt{T_2^*/T_2}$:
\begin{equation}
\eta_{AC}  
\sim \frac{\pi \hbar}{2 g \mu_B C \sqrt{T_2}}.
\label{Eq9}
\end{equation}

The optimum sensitivity is achieved only for fields oscillating near $\nu \sim 1/T_2$.  However, these results can be easily extended to higher frequency signals.  In particular, for signal field oscillation periods shorter than the dephasing time,
the interrogation time need not be restricted to the duration
of one period, but can be multiples of it.  Then, composite pulse sequences such as CPMG\cite{CPMG} may perform better
at the expense of a reduced bandwidth.  Furthermore, in ultra-pure samples where nuclear spins' evolution leads to decay of the echo signal, the long correlation time of the nuclei leads to non-exponential decay of the echo signal\cite{russianBook,JeroTheory}.  In this case, the CPMG sequence can increase the interrogation time, further reducing the minimum detectable field (see Fig.~2 and Methods).
Finally, another way to improve the magnetometer sensitivity is to use many sensing spins, where we can take advantage of the relatively high achievable density of spins in the solid-state ($\sim 10^{17}$ cm$^{-3}$) compared to atomic magnetometers($\sim 10^{13}$ cm$^{-3}$).\cite{collisionNoise}  

\section*{Implementation with NV centers}

\begin{figure}
	\centering
\includegraphics[width=4in]{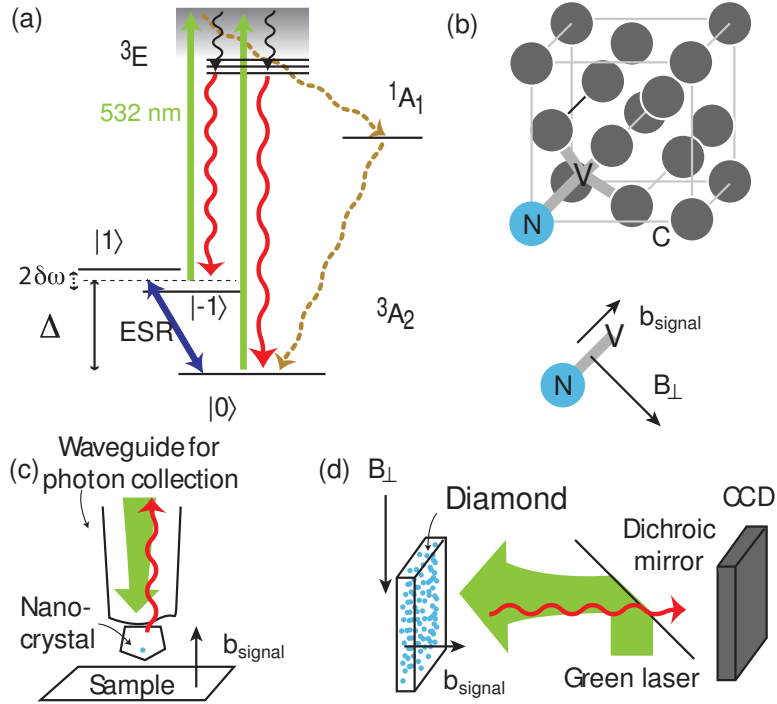}
\caption{Overview of a diamond-based magnetometer.  
(a) Level  structure of a single NV center. The NV center's ground state is a
spin triplet with a $\Delta = 2.87$ GHz crystal field splitting and
a Zeeman shift $\delta \omega$.  Under the application of green
light ($\sim 532$ nm), the NV center initially exhibits spin-dependent
photo-luminescence, even
at room temperature, allowing for optical detection of
electronic spin resonance. After continued illumination the NV spin is pumped into the ground state $m_s=0$.  
(b) Crystal structure of diamond with a (111) NV center.  
A static bias field $B_\perp$ is applied perpendicular to the 111 axis,
while small magnetic fields aligned with the 111 axis are detected as the
signal.  
(c) A nano-crystal of diamond at the end of a waveguide for photon collection, 
with resolution limited by the size of the crystal, or 
(d) a  macroscopic sample of diamond, with resolution limited by optics, allows for high spatial resolution and
signal-to-noise.  A green laser produces spin-dependent photo
luminescence, detected by measuring red light imaged onto a CCD.}
\label{f:schematic}
\end{figure}

We now discuss specific details of magnetometry using Nitrogen-Vacancy centers in diamond, developing an appropriate operating regime and 
determining the optimal sensitivities possible for current experimental technology.
The fine structure of
the electronic ground state of a NV center, shown in
Fig.~\ref{f:schematic}-a, is a spin triplet.  The crystal field splits
the $m_s=\pm 1$ Zeeman sublevels from the $m_s = 0$ sublevel by
$\Delta = 2 \pi \times 2.87$ GHz,  allowing the use of
electron-spin resonance (ESR) techniques even at vanishing external
magnetic field.  Furthermore, under application of green light, NV centers exhibit a transient, spin-dependent fluorescence, which allows for optical detection of the spin.  After the transient signal decays, the system optically pumps into the $m_s=0$ state, which prepares the system for the next measurement (see Methods).

As a specific example, we focus on magnetometry in low  external
static magnetic fields ($\leq$ 10 mT).  In this case, $\Delta$ is the largest energy scale
and sets the spin quantization axis parallel to the
nitrogen-to-vacancy direction.  The secular Hamiltonian, including a
small external field $\vec{B}(t) = (B_x,B_y,B_z)$, is
\begin{equation}
\ham = \hbar \Delta S_z^2 + g \mu_B B_z S_z,
\label{Eq1}
\end{equation}
where $B_z$ is the component of the magnetic field along the NV
center's axis and $S_z$ takes the values $m_s = 0,\pm 1$.  Terms
proportional to the perpendicular field are suppressed to order $\sim
B_{x,y}^2/\Delta$ and do not depend on the field $B_z$ being measured, and therefore may be neglected.

At low magnetic fields, the $m_s=\pm 1$ manifold can
be used to implement a vector magnetometer, sensitive only to components of the magnetic field along the center's axis. 
We focus on the $m_s=\pm 1$ manifold as it has twice the
energy splitting of the 0-1 manifold and is less affected by nuclear
spin-induced decoherence at low fields, since inter-nuclear
interactions are suppressed by the large hyperfine
field\cite{diffusionbarrier}.

Coherent control of the NV center's spin states is obtained via an ESR
magnetic field oscillating at angular frequency $\Delta$.  ESR
pulses linearly polarized along the x-axis rotate the NV spin between the two
dimensional subspace of $\ket{0}$ and $\ket{+} = (\ket{1} +
\ket{-1})/\sqrt{2}$.  To manipulate $\ket{\pm1}$ superpositions,
additional control can be provided by a background oscillating
reference field $[B_{\rm ref} \sin(2 \pi t/\tau)]$ along the $z$-axis.
Specifically, $B_{\rm ref} = (\hbar/g \mu_B) \pi^2 / 8 \tau$ yields an
optimal phase offset to achieve a magnetometer signal linear in the
field strength (Fig. 2).  

The sensitivity as a function of the signal
frequency for both AC and DC detection is plotted in
Fig.~\ref{f:sens}.  For diamond where natural abundance (1.1\%)
Carbon-13 nuclei are the principal cause of dephasing, $T_2^* \sim 1
\mu$s and $T_2 \sim 300 \mu$s\cite{dutt06}.  
Again using current experimental parameters, with $C \approx 0.05$, 
and measurement and preparation
time $t_m \leq 2\ \mu$s, we can optimize the sensitivity as a function of $\tau$.  Including corrections from decoherence with expected signal
decay (see Methods) $\propto \exp[-(\tau/T_2)^3]$
we find:
$\eta_{AC} = \frac{\pi \hbar}{2 g \mu_B} e^{(\tau/T_2)^3} \sqrt{\tau + t_m} / C \tau$.
We obtain optimal sensitivity of $\eta_{AC} \approx 18$ nT Hz$^{-1/2}$
for a single NV center using current experimental collection efficiencies.  Improved collection efficiencies ($C=0.3$) would yield $\eta_{AC} = 3$ nT/Hz$^{-1/2}$.
Note that spin $T_1$ relaxation occurs on
timescales much longer than milliseconds and may be safely neglected\cite{dutt06}.  Finally,
the observed dephasing times are independent of temperature from 4 K
to 300 K, due in part to the vanishing polarization of the nuclear
bath at small magnetic fields.

When more than one nitrogen vacancy center exists in the sample, 
they can belong to four different crystallographic classes, 
each corresponding to the centers' alignments along different (111) axes. 
To operate as a vector magnetometer along
a controlled direction, a transverse (DC) magnetic field
$B_{\perp} \geq 0.3$ mT (see Methods) detunes the other three classes' levels such that the ESR field used for
quantum control excites only spins with the desired crystallographic
orientation, perpendicular to the external field.  Thus 1 in 4 spins contribute to the magnetometer signal.

\section*{Magnetometry in the high density limit}

A principle advantage of our approach over other spin precession
magnetometers is the high achievable density $n$ of sensing spins.  This improves the sensitivity to fields that are homogeneous over
the magnetometer volume, since the projection-noise per unit volume
decreases as $1/\sqrt{n}$.
NV centers can be created in controlled densities
by implanting high-purity diamond with nitrogen ions and subsequently
annealing the sample to recombine the nitrogen with
vacancies\cite{NVCProd06}.  Assuming an initial nitrogen concentration
$\sim 10^{18}$ cm$^{-3}$ with a conversion $f \sim 0.1$ to NV
centers\cite{NVCProd05,NVCProd06,wracthnphys}, we 
expect it will be feasible to create
diamond crystals with an NV center density exceeding $\sim
10^{17}$ cm$^{-3}$, with an average distance between centers of less
than 10 nm.  Even at these densities, effects such as superradiance do not play a role 
due to the large spectral width of the NV fluorescence.

At high spin densities, NV-paramagnetic impurities and NV-NV interactions 
may limit the sensitivity of the magnetometer.  
In particular, substitutional Nitrogen impurities with one bound electron (P1 centers) become a
sizable source of dephasing in high density samples\cite{charnock01,ScienceAschalom}.  The dipole-dipole
interaction between these centers has a characteristic time scale
$T_c \equiv \frac{1}{\alpha n_{epr}}$, where $\alpha$ is on the order of the
dipole coupling between electron spins,
$\frac{\mu_0}{4\pi}\frac{(g\mu_B)^2}{\hbar} \approx 3.3\times10^{-13}~
\textrm{s}^{-1} \textrm{cm}^3$, and $n_{epr}$ is the density of
paramagnetic impurities.  Qualitatively, this time scale corresponds to
the rotation time of a single paramagnetic spin in the presence of the
random field from the other paramagnetic  centers.  The time scale for interaction
between this impurity bath and a given NV center will be of the
same order of magnitude.  This suggests an exponential decay of spin
echo coherence on a timescale $T_c$ (see Methods), in
contrast to single NV center-based sensing 
where nearby nuclear spins
limit the coherence time.  

To evaluate the effects of paramagnetic impurities, we assume a density $n$
of NV centers and $n_{epr} = n(1-f)/f$ of paramagnetic impurities, where $f$ is the conversion factor described above.
The relevant figure of merit is the sensitivity per root volume 
$\eta_{AC}^{V}=\eta_{AC}\sqrt{V}$.  We find
\begin{equation}
	\label{sensitivityZZ}
	\eta_{AC}^{V}= \frac{\hbar}{g \mu_B} \frac{\pi e^{(\tau 
         /T_{2,Carbon})^3 }
     }{C \sqrt{n~\tau}} \times e^{\tau/T_c },
\end{equation}
where we have taken into account that the sensing centers account for only one
fourth of the NV centers in the sample. Here we include both dephasing due
to a bath of dipolar-coupled nuclear spins and the paramagnetic spin
bath just discussed.  In the high NV density- and low $f$-regime,
$T_{2,Carbon} > T_c > T_2^*$, i.e., Carbon-13 is no
longer the limit to echo lifetimes, but still limits
inhomogeneous broadening.  Then the optimum magnetometer sensitivity becomes:
$\eta_{AC}^{V}* = \frac{\hbar \pi}{g \mu_B C} \sqrt{\frac{2 \alpha e
  (1-f)}{f}}$.  For $f=0.1$ and $T_{2,Carbon} = 300\ \mu$s, the optimum sensitivity is NV density independent over the range $n \simeq 10^{15}-10^{17}$ cm$^{-3}$, as is seen in Fig.~\ref{Density}a, and reaches a maximum sensitivity value $\eta_V \sim 250$ aT Hz$^{-1/2}$ cm$^{-3/2}$ for $C=0.3$.
However, 
the optimum echo time depends upon the NV density, 
$\tau = f/[(1-f)2 n \alpha]$, with higher density samples requiring higher detection frequencies.
Finally, for $n \gg 10^{17}$ cm$^{-3}$,
corrections due to finite preparation, control, and measurement times
can become important, and lead to the limitations in sensitivity at high NV density seen in Fig.~\ref{Density}a.

To push the sensitivity limits beyond the cutoff imposed by paramagnetic impurities, we
can exploit more advanced forms of dynamical decoupling\cite{viola98}
than spin echo.  With appropriate external time-dependent controls, the
system can be made to evolve under an effective, time-averaged Hamiltonian
that is a suitable symmetrization of the undesired
interactions.  For example, driving the P1 centers via spin
resonance at a rate much faster than the intrinsic decorrelation time,
$T_c$, acts as a rapid spin-echo for the NV centers without impacting the NV center's magnetic field sensing capabilities.
Furthermore, improving implantation and conversion techniques (by
optimizing implant energies\cite{NVCProd05} or by using cold implantation\cite{NVCProd88}) could increase the ratio of NV centers to paramagnetic
impurities. When the conversion efficiency exceeds 50\%, interactions
between NV centers become the primary source of noise, with a
dephasing\cite{slichter} $\propto (\alpha n \tau)^2$.  The
coupling between the sensing NV centers is a $S_{j,z}S_{k,z}$
interaction which is not removed with spin echo.  However, by using collective rotations driven by appropriate ESR pulses,
the interaction can be successively rotated through the x, y and z
axes for an equal time duration\cite{Mehring}; so that on average the
spins will experience an isotropic Hamiltonian, which commutes with
the signal perturbation and thus allows the spin evolution necessary
for magnetometry\cite{MPMRey07}.  Pulse sequences such as MREV\cite{Mansfield,khodjasteh:062310}
can induce the desired evolution, and will be necessary in the high
NV-center density limit.

\section*{Single spin detection with a nano-NV-magnetometer}

NV magnetometers can be applied to an outstanding challenge in magnetic sensing: 
the detection
and real space imaging of small ensembles of electronic and nuclear
spins, with the long-term goal of resolving individual nuclear
spins in a molecule. Since the magnetic field from a single dipole
decreases with distance as $\sim 1/r^3$, a magnetometer that can be
brought into close proximity of the field source offers a clear
advantage. A diamond
nanocrystal or a single NV center near the surface of a bulk crystal would allow for a spatial resolution limited only by the distance between the NV center and the object of study,
not by the wavelength of the fluorescence signal.  For example, consider as a prototype system consisting of a crystal with a single NV center at a distance $r_0 \sim 10$ nm from the surface of the crystal.  
At this distance, the dipolar field from a single proton is $B_{\rm H} \simeq 3$ nT, which is well within the projected limits for a single NV center.

To examine a practical method to measure the magnetic field from a single spin, we consider a material with a varying nuclear spin density $n_s$ that is brought in close proximity (a distance $\sim r_0$) to the NV center.  At realistic temperatures, the thermal nuclear spin polarization of the material will be small.  However, because only a few spins are
involved, 
the distribution of spin configurations leads a large variance in the 
spin polarization\cite{mamin05}, providing a substantial,
albeit randomly oriented, magnetic field to be detected by the NV magnetometer.
We find (see Methods) that the field magnitude measured by our sensor will be characterized by a variance $B_{\rm rms} \sim B_{\rm H} \sqrt{N}$, where $N \sim 8 \pi n_s r_0^3$ is the effective number of spins contributing to the signal.
This indicates that our prototype system has an effective spatial resolution determined only by the distance of the NV center from the surface of the sample material, assuming one can position the sensor relative to the sample with stability much better than $r_0$.  

At nuclear spin densities $\lesssim 10^{18}$ cm$^{-3}$, there is on average one or fewer nuclear spins in an effective sensing volume with $r_0 \sim 10$ nm.  Hence, in this case single spins could be measured.
However, most organic molecules have substantially higher proton
densities ($\sim 10^{22} - 10^{23}$ cm$^{-3}$).  To measure only one proton at a time would require a further improvement in spatial
resolution.  
In this case  a  magnetic field gradient can be used, which allows one to convert high spectral resolution to high spatial resolution. Using similar techniques
to those used in magnetic force resonance microscopy\cite{mamin07}, a magnet near the surface of a
substrate can produce gradient fields of $ \gtrsim 10^6$
T/m (Fig.~\ref{f:ensemble}b). The narrow bandwidth of our detector, $\sim 4$ kHz $(\sim 1/T_2)$, allows it to spectrally distinguish 
two protons separated by a magnetic field difference of 0.1 mT, corresponding to physical separation of 0.1 nm. This implies
that individual proton detection may be possible even in organic and biological molecules. The narrow bandwidth associated in particular  with the CPMG approach (see Methods) allows one to distinguish
different isotopes, due to their unique gyromagnetic ratios.  More generally, our approach enables the detection of nanoscale variations in the chemical
and physical environment.

We note that the present approach can surpass the sensitivity of
SQUID\cite{bending99}, Hall-bar\cite{chang:Hallmag}, and recently proposed optically-pumped semiconductor-based\cite{meriles05} nano-magnetometers by more than an order of magnitude, with 10-1000 times
better spatial resolution.  The ultimate limits to miniaturization of
NV center nano-magnetometers, which are likely due to surface effects, are not
yet well understood.  

\section*{Imaging of macroscopic magnetic fields}

In contrast to the nano-magnetometer approach outlined above, a
macroscopic crystal of diamond containing many NV centers may be used as a high sensitivity
imaging magnetometer with large field-of-view and optical wavelength-limited spatial
resolution.  As an example system we consider a crystal of
diamond with a high density of NV centers.  The signal from NV centers in a diffraction limited setting, where a CCD might be used to image the crystal, is divided into separate ``pixels'', with each pixel corresponding to a $\sim (1 \mu$m$)^{3}$ volume element of the crystal. 
For NV center densities of $\sim
10^{15}-10^{17}$ cm$^{-3}$ and $C=0.3$, each pixel would have $\sim$ 100 pT Hz$^{-1/2}$
AC sensitivity.  This spatial resolution is comparable to micro-SQUID
magnetometers but with four orders of magnitude higher magnetic field sensitivity~\cite{veau2002}.   In such a scenario, diamond crystals could range from tens of microns to millimeters in size; and be physically integrated with fiber-based optics for a robust and practical magnetic field imager.
 
Larger detector volumes 
further improve the sensitivity for whole-sample measurements.
For example, a (3 mm)$^2$ x 1 mm thick
crystal can achieve an overall sensitivity of 3 fT Hz$^{-1/2}$ with mm resolution. 
Reducing the ratio of paramagnetic impurities to
NV centers could potentially lead to the detection of attotesla
fields, opening the prospect of improved tests of fundamental
symmetries and physical laws.

\section*{Conclusions}

The extremely high magnetic field sensitivity in a small volume
offered by solid state spin-qubits such as NV centers in diamond can
find a wide range of applications, from fundamental physics tests or
quantum computing applications to detection of NMR signals, surface
physics and material science, and medical imaging and biomagnetism.
Recently, a proof-of-principle experimental demonstrations of such a sensor have been performed by members of our collaboration\cite{JeroMagnetometer} and other groups\cite{StutMag}.
Further extensions could include the use of
non-classical spin states, such as squeezed states induced by the
spin-spin coupling. The sensitivity could also be improved by using
synthesized, isotopically purified diamond containing a lower
fraction of Carbon-13, the main cause of dephasing at moderate NV
densities, and by developing more efficient NV center creation
techniques that do not result in high densities of paramagnetic impurities. On a more general
level, these ideas could apply to a variety of paramagnetic systems or
other types of solid-state qubits that are sensitive to different perturbations. 

\begin{figure}
	\centering
\includegraphics[width=5in]{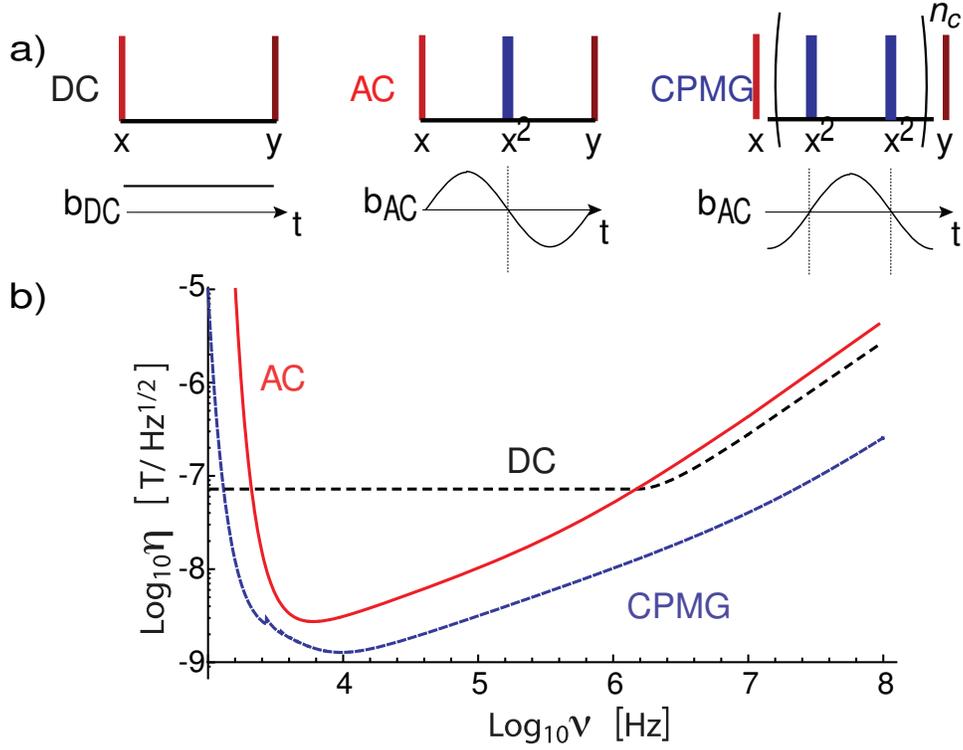}
\caption{
a)  ESR pulse sequences for magnetometry, where $x$ and $y$ indicate the linear polarization of the ESR pulse in the lab frame. 
Left: Ramsey pulse sequence  for DC-field measurement. 
Middle:  echo-based pulse sequence for AC magnetometry $\pi/2|_x-\pi|_x-\pi/2|_y$. Right: CPMG-based pulse sequence for improved AC magnetometry  \mbox{$\pi/2|_x($-$\pi|_x$---$\pi|_x$-$)^{n_c}\pi/2|_y$}, where $n_c$ is number of repetitions of the paired $\pi$ pulses.
For small accumulated phases, a signal linear in the field  can also be obtained with all pulses along the x direction if a reference field $ B_{\rm ref} \sin(2 \pi t/\tau) $ is added.
b) DC and AC sensitivity to magnetic fields 
for a single NV center as a function of
signal frequency, $\nu$.  Also shown is the expected performance of CPMG composite pulse sequences, with the optimum $n_c$ as described in Methods. 
Parameters used assume Carbon-13 limited coherence with $T_2^* = 1\ \mu$s,\cite{jelezko:130501}
$T_2 = 300\ \mu$s,\cite{childress06} $t_m = 1\ \mu$s, $C=0.3$, $T_1 = 20$ ms,\cite{dutt06} and an error per pulse of 1\%.  
}
\label{f:sens}
\end{figure}

\begin{figure}
	\centering
		\includegraphics[width=5in]{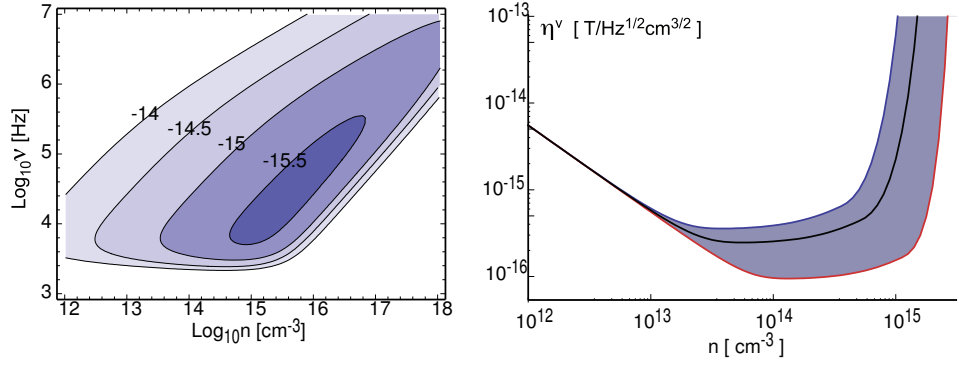}
\caption{Sensitivity-per-root-volume $(\eta^V_{AC})$ at high NV center density, for the AC-field echo measurement scheme. Left: Contour plot of $\log_{10}{\eta^V}$ in T  Hz$^{-1/2}$cm$^{3/2}$ as a function of NV center density and signal field frequency. Right: sensitivity at the optimal field frequency, as a function of NV center density; the black curve is the sensitivity for $f=.1$ while the blue and red curve are for $f=.05$ (higher paramagnetic impurity density) and $.5$ (lower paramagnetic impurity density), respectively. 
Parameters used correspond to $T_{2,Carbon} = 300\ \mu$s,\cite{childress06}, $t_m = 1\ \mu$s, $C=0.3$ and $\alpha=3.3\times10^{-13}~\textrm{s}^{-1} \textrm{cm}^3$.
}	\label{Density}
\end{figure}

\begin{methods}
\subsection{ESR control techniques}
The NV center's spin triplet has a V-type level configuration.  An
external microwave field tuned to the $\Delta = 2.87$ GHz resonance
with its magnetic field linearly polarized along the x-axis drives
transitions between $\ket{0}$ and the superposition $\ket{+}=(\ket{1}
+ \ket{-1})/\sqrt{2}$, while the state $\ket{-}=(\ket{1} -
\ket{-1})/\sqrt{2}$ is dark---it is decoupled from the field due to
destructive quantum interference.  Application of a magnetic field
aligned with the NV-center $z$-axis perturbs the interference, and
allows for complete quantum control of the spin triplet.  In
an echo sequence  appropriate for magnetometry using the $\ket{+}$ and $\ket{-}$ states, the traditional $\pi/2- \pi - \pi/2$ structure is
replaced by $\pi - 2 \pi - \pi$: the first pulse creates $\ket{+} $,
the second induces a relative $\pi$-phase shift between $\ket{+}$ and
$\ket{-}$, and the third converts $\ket{+}$ to $\ket{0}$ while leaving
$\ket{-}$ population trapped in the $m_s = \pm 1$ manifold.  We remark that
for external fields in excess of a few mT it may be more convenient to use
the 0-1 manifold, as two different resonance frequencies would be necessary for using the $\pm 1$ manifold in this regime.

\subsection{AC-field measurement scheme and bandwidth}
AC-field detection requires synchronization of the pulse sequence with the external magnetic field oscillations. When this is not practical or if the field phase $\varphi_0$ varies randomly in time, successive measurements will give random readings distributed over the range of the function $f(\nu \tau,\varphi_0)$ (given in the main text) resulting in a zero average signal. 
In this situation, information about the field intensity is contained in the measured signal variance, provided the random phase correlation time $\tau_\varphi$ satisfies: $\tau\ll\tau_\varphi <T$. (If $\tau_\varphi >T$, the total averaging time, the  scheme presented in the main text could be used). For $\tau= 2 \pi/\nu$ and  a uniformly distributed $\varphi_0$, $\langle f(2\pi,\varphi_0)^2\rangle=2/\pi^2$ and the standard deviation of the measured signal is: $\frac{g \mu_B \sqrt{2}}{\hbar\pi} b\tau$, while the noise has a contribution from the uncertainty in the variance equal to $2^{1/4}/\pi$.  The sensitivity is thus only worsened by a factor  $\sqrt{2(1+\sqrt{2}/\pi^2)}\approx 1.5$ compared to detection of a signal with a known phase.

To increase the sensitivity at higher frequencies, it is possible to increase the interrogation time (see main text) by using a series of $2\pi$-pulse cycles (CPMG pulse sequence).  A single cycle corresponds to the pulse sequence $\tau/4 - \pi - \tau/2 - \pi -\tau/4$. While this method increases the sensitivity, the measurement bandwidth  decreases with increasing cycle number $n_c$. The AC magnetometer response to a general signal $b(t)$ can be calculated from a frequency space analysis: 
$\frac{1}{2\pi}\int_{-\infty}^\infty {\tilde{b}(\omega)\left(\int_0^{\tau/2}{e^{i\omega t} dt}-\int_{\tau/2}^\tau {e^{i\omega t} dt}\right)d\omega}=\frac{\tau}{2\pi}\int_{-\infty}^\infty {\tilde{b}(\omega)W_0(\omega,\tau)e^{i\omega \tau/2} d\omega}$, where $\tilde{b}(\omega)$ is the Fourier transform of the signal field and $W_0$ a windowing function. With a similar calculation we obtain the windowing function for an $n_c$-cycle pulse sequence: 
\begin{equation*}
W_{n_c}(\omega,\tau) = \frac{1 - \sec(\tau \omega/4)}{\tau \omega/2} \sin(n_c \tau \omega/2)
\ .
\end{equation*} 
This function has
a band-center $\approx 2 \pi/\tau$ and bandwidth (HWHM) $\sim  4/(n_c\tau)$.

We can evaluate the improvement in coherence times for the CPMG sequence in cases where a detailed understanding of the main source of decoherence is available.  For the single-spin magnetometer, we can approximate the nuclear spin environment by separating contributions from distant nuclear spins, undergoing dipolar spin diffusion, and nearby nuclear spins, whose evolution is frozen by the electron spin's dipolar field\cite{russianBook,JeroTheory}.  We can model the distant nuclear spins as an exponentially correlated gaussian fluctuating field $\tilde{B}$ with a correlation function $\mean{\tilde{B}(t)\tilde{B}(t')} \sim (\frac{\hbar}{g \mu_B T_2^*})^2 \exp(-|t-t'|/T_c)$ where $T_c \gg T_2^*$ is the correlation time of the nuclear spins.

Within this model, the random phase accumulated during an echo sequence ($\delta\phi = \frac{g \mu_B}{\hbar} \int_0^{\tau/2} \tilde{B}(t) dt - \int_{\tau/2}^\tau \tilde{B}(t) dt$) is characterized by its variance, $\mean{\delta\phi^2} 
\approx \tau^3/[6 T_c (T_2^*)^2]$ for $T_c \gg T_2^*,\tau$.  Applying this model to an $n_c$-cycle CPMG sequence gives $\mean{\delta \phi^2} \sim (n_c \tau)^3/[24 n_c^2 T_c (T_2^*)^2]$. 
Thus the multiple-pulse sequence yields an improvement in the lifetime by $(2 n_c)^{2/3}$.\cite{slichter}  The improvement is conditional on $\tau, T_2^* \ll T_c$ and on
the total interrogation time $n_c \tau$ being less than the relaxation time of the electron spins.  Recent experiments have shown that the relaxation time in ultra-pure samples is $\gg 20$ ms\cite{dutt06}, suggesting $n_c \gtrsim 40$ cycles can result in an $(2 n_c)^{1/3} \gtrsim 4$ overall improvement in sensitivity.  Note that in practice this improvement will be limited by imperfections in the control pulses.  For example, $\pi$-pulse errors of order 1\% will limit $n_c \approx 25$, resulting in the optimal sensitivity show in Fig.~\ref{f:sens}.

\subsection{Measurement efficiency}
The state of the electronic spin is measured by spin-selective
fluorescence. When illuminated by green light, NV centers in the
$m_s=0$ state undergo a cyclic transition\cite{Manson}, with a rate
limited by radiative decay ($\gamma \sim 15$ MHz).  At the same time,
centers in the $m_s=\pm 1$ state are rapidly pumped into a dark
singlet state, from which they decay to the $m_s=0$ state after a time
$t_p\approx 0.5\ \mu$s. To allow for a good discrimination of the
$m_s=0,\ \pm1$ states, the measurement time $t_m$ should be smaller
than the optical pumping time $t_p$.

For a given photon collection efficiency $\eta_m$, an average of
$\alpha_0 \simeq (t_m \gamma) \eta_m$ photons are detected from each
spin in the $m_s=0$ state and $\alpha_1 (< \alpha_0)$ photons are detected
from each spin in the $m_s=\pm 1$ manifold.  We can estimate the
combined effects of spin projection-noise and photon shot noise for $N$
measurements as $N^{-1/2} /C$, recovering the formulae for sensitivity
used in the main text, with $1/C = \sqrt{1 +
2\frac{\alpha_0+\alpha_1}{(\alpha_0-\alpha_1)^2}}$.  This includes
the effects of photon shot noise and reduced contrast.  For current
experiments, a contrast $(\alpha_0 - \alpha_1)/(\alpha_0 + \alpha_1)
\sim 0.3$ is observed.  Efficiencies of $\eta_m \sim 0.001$ are achieved in current experiments\cite{childress06,dutt06} and give $C \sim 0.05$.   Assuming 
high collection efficiency ($\eta_m \gtrsim 0.05$) gives
$C \sim 0.3$.

\subsection{Effects of different NV center orientations}
In order to use an ensemble of NV centers as a vector magnetometer, the signal
should originate only from one of the four different crystallographic
axes.  Under application of a DC transverse
magnetic field $B_{\perp} \hat{x}$, the other (spectator) centers (with
crystalline axis $\hat{n}$) have their $\ket{\pm 1}$ levels split by $g
\mu_B B_{\perp} \hat{x} \cdot \hat{n}$.  This detunes the spectator centers from the microwave field used for preparing and
manipulating the $m_s = \pm 1$ subspace.
For example, to use NV centers along the (1,1,1) crystallographic axis, the ideal choice of
$\hat{x}$ is to align it with the ($1,1,\bar{2}$) axis.  We require the
microwave Rabi frequency $\Omega \geq 3 \pi/T_2^*$ for pulse errors to
be smaller than our assumed measurement errors for the desired (111)
axis.  This translates to a requirement that $g \mu_B B_{\perp} > 3
\hbar \Omega \sqrt{3/2}$ for the other three axes.  For $T_2^* = 1
\mu$s, we require $B_{\perp} \geq 0.3$ mT.  
One intriguing development of NV center based magnetometry would be to
exploit the four crystallographic classes of NV centers to provide a
full (3D) vector magnetometer, by changing the direction of
the biasing transverse field $B_\perp$ in between measurements.

Errors due to inhomogeneities in the NV center properties
(e.g., variations of the $g$-factor due to crystal strains) or to spatial
inhomogeneities of the magnetic field can typically be neglected. Even for
an average microtesla signal field, a distribution of $g$-factors of field inhomogeneity of order induced dephasing is 4\% leads to a broadening of the signal that is smaller than the effects of $T_2$.

\subsection{Coupling to paramagnetic impurities}
The coupling of an NV electronic spin to other NV centers
($\vec{S}_k$) and paramagnetic (epr) impurities such as nitrogen
($\vec{I}_k$, $g_I\approx g$) is given by the magnetic dipolar
interaction.  To first order in $1/\Delta$, the secular dipolar
Hamiltonian is given by: $\ham_{zz} + \ham_{epr}$, with $\ham_{zz} =
\sum_{jk} S_{z,j} \vec{D}_{jk} \cdot \hat{z}_k S_{z,k}$ and
$\ham_{epr} = \sum_{jk} S_{z,j} \vec{D}_{jk} \cdot \vec{I}_{k}$.  The
dipole interaction vector is
$\vec{D}_{jk}=\frac{\mu_0g^2\mu_B^2}{4\pi\hbar}\frac{[3(\hat{r}_{jk}
\cdot \hat{z}) \hat{r}_{jk}- \hat{z}]}{r_{jk}^3}$, with the
$\hat{z}$ axis set by the N-V crystal axis of the sensing spin
centers.

We model the secular component of the dipole coupling between
paramagnetic impurities and NV centers as $\omega_{jk} = \vec{D}_{jk}
\cdot \hat{x} I^k_x$ to the $j$th NV center, and with a characteristic
correlation time $t_c \approx \hbar/\sqrt{\langle D^2 \rangle}$:
$\mean{\hat{\omega}_{jk}(t) \hat{\omega}_{jk}(t')} \approx \langle D^2
\rangle \exp(-|t-t'|/T_c)$.  We can now calculate the expected
spin-echo signal as a function of $\mean{D^2}$, which scales as the
square of the density of paramagnetic impurities.  In this limit, when
the correlation time and the interaction energy are at comparable
scales, spin echoes decay exponentially as $\exp (-t/T_c)$.  We find
$T_c \approx 4/\sqrt{\alpha n^2}$; hence for paramagnetic impurity densities of
$10^{19}$ cm$^{-3}$, $T_c \approx 1\ \mu$s.

At high densities, paramagnetic impurities and spectator NV centers may
have sufficiently strong interactions to reduce the correlation time
of the field-aligned component, $I_x$.  Spectator NV centers may be
optically pumped to their $m_s=0$ state, reducing dynamical noise reducing the effective temperature of the spectator system.  However, spin echoes will
not remove the effects of the paramagnetic impurities with short
correlation times, and they may in fact limit the $T_2$ time and the
corresponding bandwidth of the system.  Experiments in systems with
high nitrogen concentrations indicate exponential decay of echoes on a
$5-10\ \mu$s timescale\cite{manson88,charnock01} due to this
coupling; more generally, the decay scales with the density of
impurities.  While approaches such as CPMG and more complex decoupling
may help, we anticipate that paramagnetic impurities concentrations
below $10^{18}$ cm$^{-3}$ will be necessary to achieve the best
predicted sensitivities of this paper.

\subsection{Number of spins detected by a point-like sensor}

To estimate the number of spins that a localized sensor will detect we determine the maximum and root-mean-square magnetic fields from a randomly distributed set of dipolar spins.  We denote the dipolar field at a position $\vec{r}_0$ from a spin $i$ as $\mathcal{G} \vec{b}(\vec{r}_i-\vec{r}_0,\vec{I}_i)$, with normalization of $\mathcal{G}=\frac{g \mu_n \mu_0}{4 \pi} = B_{\rm H} (10\ {\rm nm})^3$ for protons and $\vec{b}(\vec{r},\vec{I}) = \frac{1}{r^3}(\vec{I}-3 \vec{r}(\vec{r} \cdot \vec{I})/r^2)$ being the position dependence of the dipolar field ($B_{\rm H}$ is the magnitude of magnetic field created by a proton at a distance of 10 nm).  The maximum detectable field occurs for polarized spins pointing perpendicular ($z$ axis) to the surface (Fig.~\ref{f:ensemble}a).  By symmetry, this field is parallel to the polarization, and we find
\begin{eqnarray*}
B_{\rm max} & = & \mathcal{G} \left\langle\sum_i  b_z(\vec{r}_i,I \hat{z}) \right\rangle_{pos} \\
& = & - 2 \pi  \mathcal{G} I  n_s\ ,
\end{eqnarray*}
where we choose coordinates such that the half-plane begins at $z = -r_0$, $\mean{\ }_{pos}$ averages over a homogenous distribution of spin positions, and we assume a density $n_s$ of dipolar spins, allowing us to replace the sum $\sum_i$ with an integral $n_s \int_{z<-r_0} d^3r$.
At high temperatures, the fluctuations of the potential values of the dipolar field reflect the $\sqrt{N}$ noise statistics from a set of $N$ spins.  The mean-square of the $z$-component of the magnetic field is then:
\begin{eqnarray*}
B_{\rm rms}^2 & = & \mathcal{G}^2 \left\langle\sum_{ij} \left\langle b_z(\vec{r}_i,\vec{I}_i) b_z(\vec{r}_j,\vec{I}_j) \right\rangle_{cfg} \right\rangle_{pos} \\
& = &  \mathcal{G}^2 \frac{I(I+1)}{3} \left\langle \sum_i \frac{1}{r_i^6} \left(1 + 3 \frac{z_i^2}{r_i^2}\right)\right\rangle_{pos} \\
& = & \mathcal{G}^2 \frac{I(I+1)}{3} n_s \frac{\pi}{2 r_0^3}\ ,
\end{eqnarray*}
where the average over spin configurations at high temperature uses $\mean{I_{i,\mu}I_{j,\nu}}_{cfg}=\delta_{\mu \nu}\delta_{ij} \frac{I(I+1)}{3}$.  We find in particular that the statistical fluctuations are consistent with $B_{\rm rms} \sim B_H \sqrt{N}$, where $N \sim n_s r_0^3$.  More specifically, the effective number of spins detected $N$ can be estimated from the relation $B_{\rm rms} = |B_{\rm max}|/\sqrt{N}$.  Thus, $N = (|B_{\rm max}|/B_{\rm rms})^2 = \frac{I^2}{I(I+1)/3} (8 \pi n_s r_0^3)$.  For $I = 1/2$ this reduces to $N = 8 \pi n_s r_0^3$, equivalent in effective detection volume to a half-sphere of radius $2.29 r_0$.

\end{methods}

\begin{addendum}
\item We gratefully acknowledge conversations with  D. Awschalom,  A. Cohen,  
J. Doyle, G. Dutt, J. Maze, E. Togan, P. Stanwix, J. Hodges, S. Hong, and M. P. Ledbetter. This work was supported by the NSF,  and David and Lucile Packard Foundation.
JMT is supported by the Pappalardo Fellowship; PC is supported by the ITAMP Fellowship. \\
$^*$ These authors contributed equally.
\item[Competing Interests] The authors declare that they have no competing financial interests.
\item[Correspondence] Correspondence should be addressed to M.D.L. (email: lukin@physics.harvard.edu).
\end{addendum}

\begin{figure}
\centering
\includegraphics[width=5in]{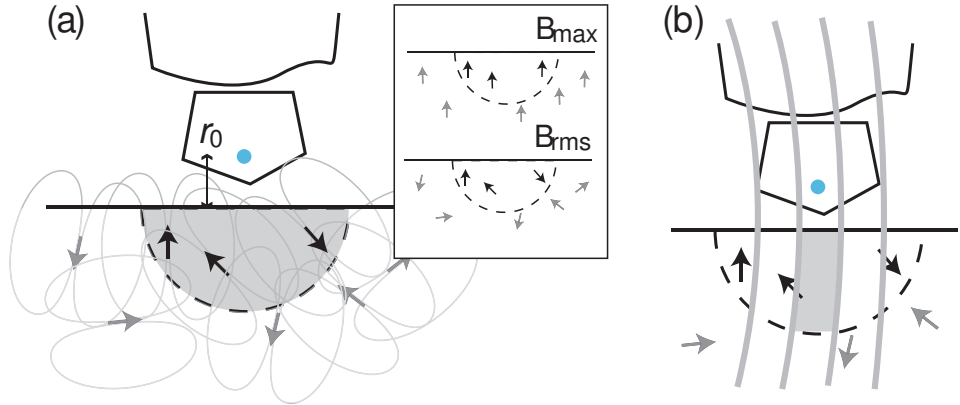}
\caption{
Illustration of high spatial resolution magnetometry with a diamond nanocrystal.  (a) The dipolar fields from spins in the sample decay rapidly with distance; only those within a distance $\sim r_o$ contribute to the observable signal for a point-like detector (such as a single NV center in a nanocrystal, illustrated by the blue dot).  The inset shows how $B_{\rm max}$ and $B_{\rm rms}$ are related; when few spins are involved, the statistical fluctuations become large.  (b) In the presence of a magnetic field gradient (field lines in gray) only a small region of the detection volume is precessing at the frequency band-center of the detector, allowing for even higher spatial resolution.
\label{f:ensemble}}
\end{figure}

\end{document}